\documentclass[aps,prx,twocolumn,superscriptaddress,longbibliography]{revtex4-2}
\usepackage{amsmath,amssymb,amsfonts}
\usepackage{graphicx}
\graphicspath{{figs/}}
\usepackage{bm}
\usepackage{xcolor}
\usepackage{braket}
\usepackage{algorithm}
\usepackage{algpseudocode}
\usepackage{booktabs}
\usepackage{dsfont}
\usepackage{wasysym}
\usepackage{bbm}
\usepackage{acro}
\DeclareAcronym{VMC}{
  short = VMC,
  long  = variational Monte Carlo
}
\DeclareAcronym{NQS}{
  short = NQS,
  long  = neural quantum state
}
\DeclareAcronym{tVMC}{
  short = tVMC,
  long  = time-dependent variational Monte Carlo
}
\DeclareAcronym{rot-tVMC}{
  short = rot-tVMC,
  long  = rotation-sampling tVMC
}
\DeclareAcronym{rot-p-tVMC}{
  short = rot-p-tVMC,
  long  = rotation-sampling p-tVMC
}
\DeclareAcronym{p-tVMC}{
  short = p-tVMC,
  long  = projected time-dependent variational Monte Carlo
}
\DeclareAcronym{TDVP}{
  short = TDVP,
  long  = time-dependent variational principle
}
\DeclareAcronym{QGT}{
  short = QGT,
  long  = quantum geometric tensor
}
\DeclareAcronym{DSF}{
  short = DSF,
  long  = dynamical structure factor
}
\DeclareAcronym{LPE}{
  short = LPE,
  long  = linear product expansion
}
\DeclareAcronym{MPS}{
  short = MPS,
  long  = matrix product state
}
\DeclareAcronym{SR}{
  short = SR,
  long  = stochastic reconfiguration
}
\DeclareAcronym{RBM}{
  short = RBM,
  long  = restricted Boltzmann machine
}
\usepackage{hyperref}

\begin{document}

\title{Resolving support-mismatch by local basis rotation in variational Monte Carlo}

\author{Jia-Lin Chen}
\affiliation{Beijing National Laboratory for Condensed Matter Physics and Institute of Physics,
Chinese Academy of Sciences, Beijing 100190, China.}
\affiliation{School of Physical Sciences, University of Chinese Academy of Sciences, Beijing 100049, China.}

\author{Zhen Fan}
\affiliation{Beijing National Laboratory for Condensed Matter Physics and Institute of Physics,
Chinese Academy of Sciences, Beijing 100190, China.}
\affiliation{School of Physical Sciences, University of Chinese Academy of Sciences, Beijing 100049, China.}

\author{Canhui Yan}
\affiliation{School of Physics and Technology, Wuhan University, Wuhan, Hubei 430072, China.}

\author{Yantao Wu}
\email{yantaow@iphy.ac.cn}
\affiliation{Beijing National Laboratory for Condensed Matter Physics and Institute of Physics,
Chinese Academy of Sciences, Beijing 100190, China.}

\author{Tao Xiang}
\email{txiang@iphy.ac.cn}
\affiliation{Beijing National Laboratory for Condensed Matter Physics and Institute of Physics, Chinese Academy of Sciences, Beijing 100190, China.}
\affiliation{School of Physical Sciences, University of Chinese Academy of Sciences, Beijing 100049, China.}

\date{\today}

\begin{abstract}
Real-time dynamics after a local quench by a charged operator encodes the response functions measured in spectroscopic experiments, yet they have long posed a challenge for variational Monte Carlo calculations.
The obstacle is a support mismatch: the projective action by a charged local operator forces an exponentially large number of configurations to vanish, but these configurations may still contribute to the dynamics, biasing the estimators and freezing the evolution at the very first step.
This difficulty is an artifact of the chosen sampling basis, and the support mismatch generated by a charged local operator is itself local. 
We demonstrate that the missing support can be restored by a local rotation of the sampling basis, without changing the underlying variational dynamics. We propose a local basis-rotation sampling scheme that resolves the support-mismatch problem and can be readily incorporated into existing variational Monte Carlo algorithms. 
Benchmarks show that rotation sampling accurately captures long-time quantum dynamics, enabling variational Monte Carlo calculations of dynamical structure factors in one dimension and unbiased local-operator quench dynamics in two dimensions.
We also show that this resolution of the support-mismatch problem extends beyond real-time dynamics, and may also be helpful for ground state variational Monte Carlo calculations.
\end{abstract}
\maketitle

\section{Introduction}\label{sec:intro}

Understanding dynamical evolution is central to the entire landscape of quantum many-body physics, quantum chemistry, and quantum information science. For large quantum systems, exact real-time evolution is impossible because the Hilbert space grows exponentially with system size. Time-dependent variational Monte Carlo~(tVMC)~\cite{tVMCCarleo2012,Carleo2014LightCone} provides a scalable alternative by combining expressive variational wave functions with stochastic estimation, which have been widely used in quantum dynamics simulations~\cite{wu2026realtimedynamicsdimensionstensor,Carleo2017,Jannes2024RealTimeDynTFD,Tiago2023tVMCSpectralFunction,Markus2020PRL10x10,Matija2023Rotor2DPinv,Gutierrez2022,Donatella2023,carleo2017unitaryBose,schmitt2018classicalNetworks,czischek2018quenchesIsing,verdel2021variationalClassical,burau2021unitaryQRG,lin2022scalingNQS,hofmann2022stochasticNoise,vandewalle2025tNQS,hartmann2019dissipativeNQS,nagy2019variationalOpen,reh2021tdvpOpenNQS}.
Instead of representing the full many-body state explicitly, tVMC evolves a compact parameterization of the wave function, while expectation values and dynamical quantities are evaluated from Monte Carlo samples.

In tVMC, the Schrödinger equation is projected to the variational manifold through \ac{TDVP}, reducing the real-time evolution to equations of motion of the variational parameters~\cite{haegeman2011TDVP}. These equations are governed by the \ac{QGT} and the force vector, both of which are estimated from Monte Carlo sampling of the quantum state~\cite{Sorella1998SR1,Sorella2001SR2,Sorella2005SR3}. However, unbiased estimation requires a support condition: the sampling distribution must cover every configuration that contributes to them.
Otherwise, a structural bias would arise where configurations with zero sampling probability are never visited and their contributions are permanently lost, leading to the support mismatch~\cite{ptvmc1Sinibaldi2023,Wan2026BlurredSampling}. 

Support mismatch is not a rare phenomenon.
It arises when the variational wave function has vanishing or negligible amplitude on physically relevant configurations.
In quantum chemistry, electronic systems are commonly formulated in real space, where anti-symmetry constraints naturally lead to nodal structures~\cite{Ceperley1991FermionNodes}.
In strongly correlated systems, the ground state typically exhibits a complex sign structure~\cite{liang2018FrustratedSignCNN,Choo20192DJ1J2NQS,Westerhout2020GeneralizationPropertiesofNeuralNetwork,Chen2022PRRSolvingSignStructure,viteritti2026approachingthermodynamiclimitneuralnetwork}, and variational optimization may require amplitudes to cross zero in order to learn this nontrivial structure.
In quantum circuits simulations, projective gates or measurements can set the amplitudes of projected-out configurations to zero~\cite{Nielsen_Chuang_2010,ptvmc1Sinibaldi2023}.

In real-time evolution, the support mismatch is a severe pathology, since once the relevant configurations lose support, the evolution becomes biased. A biased update at a single time step is non-local in time and contaminates all subsequent steps. Several approaches~\cite{ptvmc1Sinibaldi2023,Jannes2024RealTimeDynTFD,ptvmc2Gravina2025,Wan2026BlurredSampling,krinitsin2026timedependentvariationalmontecarlo} have been proposed to address this problem. 
However, these studies have primarily focused on global quenches, where the mismatch builds up gradually over the course of the evolution. Local-operator quenches have received little attention in this context, despite their relevance to experimental spectroscopy: the unequal-time correlations that follow them are precisely the response functions probed by neutron scattering and angle-resolved photoemission spectroscopy~\cite{Tiago2023tVMCSpectralFunction, villa2020LocalQuenchSpectroscopy, White2008, wang2019dynamicstructurefactorreal, Ferrari2018}.

Meanwhile, charged local-operator quenches represent one of the most explicit cases of support mismatch. For example, in a spin-$1/2$ Hamiltonian with $U(1)$ symmetry, a charged local operator such as spin raising operator $S^+$ fixes the quench-site spin to be up, so configurations with the opposite spin at the quench-site have exactly zero amplitude. This removes a large part of configuration space from the distribution, causing the simulation to fail at the very first step and putting the spectral functions accessible through tVMC out of reach.

However, these zeros are not an intrinsic property of the quantum state---they depend on the chosen basis. 
A state that has exact zeros in one computational basis may be fully supported in another. 
Such an example arises in finite-temperature VMC simulations, where a basis rotation in the auxiliary thermofield degrees of freedom redistributes the support of the infinite-temperature state, eliminating exact zeros and enabling stable imaginary-time evolution~\cite{Jannes2024RealTimeDynTFD}.
Beyond VMC, a suitable choice of basis can mitigate the sign problem~\cite{hangleiter2020EasingQMCSign,kim2020QMCMultiOrbialHubbardSign} and disentangle a quantum state~\cite{qian2024DisentanglerDMRG,qian2025DisentanglerTDVP,fan2025DisentanglerIsing}.
More broadly, basis choice can serve as a computational resource, as illustrated by randomized measurements~\cite{Elben2023RandomizedMeasureToolBox} and classical shadows~\cite{Huang2020ClassicalShadow}.

In the following, we refer to as \textit{rotation sampling}, the sampling scheme where the sampling basis is rotated to mix the zero-amplitude configurations with nonzero ones, recovering the lost support.
For the charged local-operator quenches we consider here, the support mismatch is itself local, confined to the quench site, so it can be removed by a local rotation of the sampling basis. 
This enables unbiased time evolution without additional constraints or reweighted estimators, opening a practical route to transverse dynamical response functions that were previously inaccessible to tVMC.

The paper is organized as follows. In Sec.~\ref{sec:toy_model} we first introduce a two-site toy model that illustrates the problem and its consequences analytically. 
In Sec.~\ref{sec:unbias_tVMC} we show that rotation sampling removes the bias in tVMC, enabling the extraction of dynamical structure factors from real-time correlators. 
In Sec.~\ref{sec:enhance_ptVMC} we discuss how rotation sampling can be incorporated into \ac{p-tVMC}~\cite{ptvmc1Sinibaldi2023,ptvmc2Gravina2025} for stable two-dimensional local-operator quench simulations.
Finally, in Sec.~\ref{sec:rotation_gs} we show that the method can also accelerate ground-state optimizations. We conclude in Sec.~\ref{sec:discussion} with a discussion of future perspectives.

\begin{figure}[tb]
    \centering
    \includegraphics[]{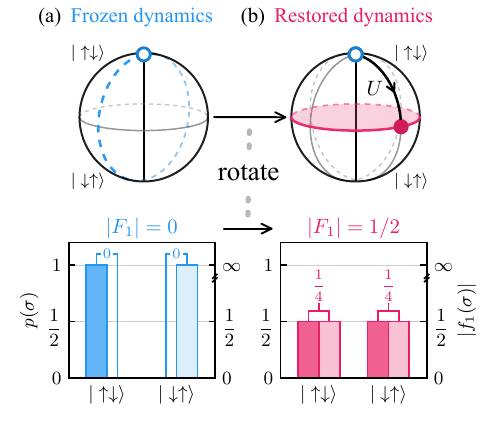}

    \caption{Schematic illustration of dynamical freezing and its resolution via rotation sampling.
        (a) Bloch-sphere representation of the dynamics within the $S^z_{\mathrm{tot}}=0$ subspace for a two-site XY model initialized in the state $\lvert{\uparrow\downarrow}\rangle$. The bar chart displays the sampling probabilities $p(\sigma)$ and the corresponding force magnitudes $\lvert f_1(\sigma)\rvert$.
        (b) Same as (a) but after applying a basis rotation $U$.
    }
\label{fig:dynamical_freezing}
\end{figure}

\section{The Support-Mismatch Problem}\label{sec:toy_model}

We start by illustrating support mismatch with a minimal two-site XY model
\begin{equation}
  H = -\bigl(S^x_0 S^x_1 + S^y_0 S^y_1\bigr).
\end{equation}
In the $S^z_{\rm tot}=0$ subspace, any state can be mapped onto the Bloch sphere, with the basis states $\ket{\uparrow\downarrow}$ and $\ket{\downarrow\uparrow}$ corresponding to the north and south poles [Fig.~\ref{fig:dynamical_freezing}(a), upper panel]. 
We parameterize the state as $\ket{\psi(t)} = c_0(t)\ket{\uparrow\downarrow} + c_1(t)\ket{\downarrow\uparrow}$. The equation of motion is
\begin{equation}
  \partial_t\ket{\psi(t)}
  =
  -\mathrm{i}H\ket{\psi(t)}
  =
  \frac{\mathrm{i}}{2}c_1(t)\ket{\uparrow\downarrow}
  +
  \frac{\mathrm{i}}{2}c_0(t)\ket{\downarrow\uparrow}.
\end{equation}
Starting from the north pole $\ket{\uparrow\downarrow}$ with $c_0(0) = 1$ and $c_1(0)=0$, the system undergoes periodic Rabi oscillation between the two poles.

However, a standard tVMC simulation of this time evolution fails already at the first step, leaving the state frozen at the north pole. The origin of this failure can be traced to the TDVP force along the tangent direction $|\partial_1\psi\rangle\equiv |\partial\psi\rangle/\partial c_1$.
In the TDVP framework, the right-hand side of the Schrödinger equation, $H|\psi\rangle$, is projected onto the tangent space of the variational manifold, yielding the force vector. At the initial state $|\psi(0)\rangle=|{\uparrow\downarrow}\rangle$, the corresponding force component can be evaluated exactly from its definition,
\begin{equation}
F_1\equiv \langle \partial_1\psi|H|\psi\rangle
= \langle {\downarrow\uparrow}|H|\psi\rangle 
=-\frac{1}{2}.
\end{equation}
However, when the same quantity is instead estimated by Monte Carlo sampling from the Born distribution $p(\sigma)=|\langle\sigma|\psi\rangle|^2$, it takes the form
\begin{equation}
F_1=
\sum_{\sigma}
|\langle\sigma|\psi\rangle|^2\, f_1(\sigma),
\end{equation}
with the estimator
\begin{equation}
f_1(\sigma)=
\frac{\langle\partial_1\psi|\sigma\rangle}{\langle\psi|\sigma\rangle}\,
\frac{\langle\sigma|H|\psi\rangle}{\langle\sigma|\psi\rangle}.
\end{equation}
This rewriting implicitly assumes that every configuration contributing to $F_1$ carries nonzero sampling probability. 
The assumption fails at the initial state: the only configuration 
$|{\downarrow\uparrow}\rangle$ that carries a contribution to $F_1$ has vanishing probability $p(\downarrow\uparrow)=|\langle{\downarrow\uparrow}|\psi(0)\rangle|^2$, and is therefore absent from the Monte Carlo estimation [lower panel of Fig.~\ref{fig:dynamical_freezing}(a)].
The sampling silently drops this contribution and returns zero. This spurious zero force is a sampling artifact we call support mismatch, and it is what freezes the variational dynamics at the initial state.

This failure can be removed in the example by evaluating the same update in a rotated representation. Specifically, one may use a rotation
\begin{equation}
 U\ket{\uparrow\downarrow}
 =
 \frac{1}{\sqrt{2}}
 \left(\ket{\uparrow\downarrow} + \ket{\downarrow\uparrow}\right)
\end{equation}
to move the initial point to the equator in the Bloch-sphere picture [upper panel of Fig.~\ref{fig:dynamical_freezing}(b)]. In this rotated representation, both configurations acquire finite sampling weight allowing the contribution that is absent in the original computational basis to enter the Monte Carlo estimate [lower panel of Fig.~\ref{fig:dynamical_freezing}(b)]. The sampled quantities are then rotated back to the original basis. Thus, for this minimal example, the rotated representation restores the missing contribution and  eliminates the dynamical freezing.

\begin{figure}[tb]
    \centering
    \includegraphics[]{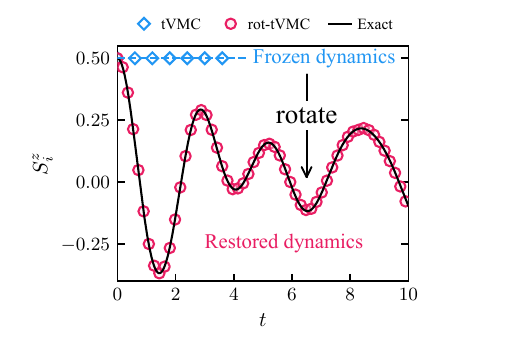}
    \caption{Time evolution of the quench-site magnetization $S^z_i(t)$ for a $4\times4$ periodic square-lattice XY model. Standard tVMC remains frozen at the initial value, whereas rot-tVMC follows the exact reference. Both simulations use an $\alpha=4$ RBM ansatz. 
    }
\label{fig:4x4_tVMC_benchamrk}
\end{figure}

\begin{figure*}[tb]
  \centering
    \includegraphics[width=0.98\textwidth]{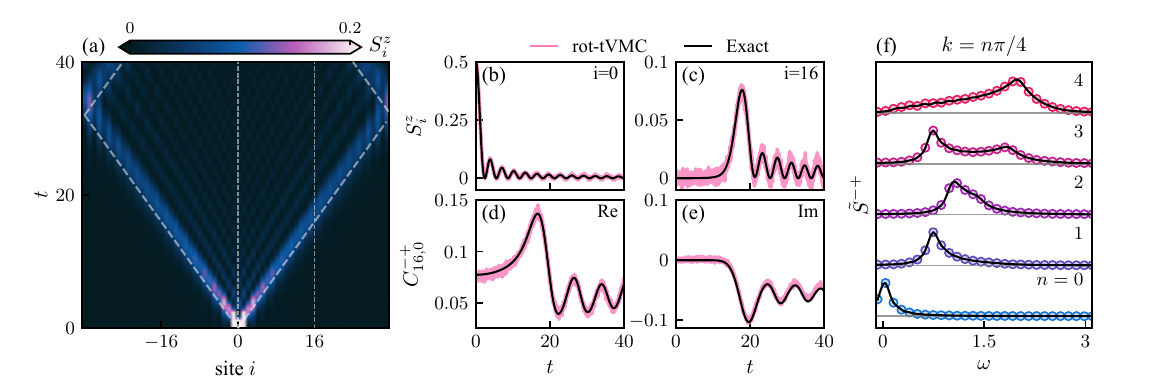}
    \caption{Local-operator-quench dynamics and dynamical structure factor of the $N=64$ periodic XY chain.
    (a) Space--time heatmap of $S^z_i(t)$ following a local quench $\ket{\psi_0} = S^+_0 \ket{\psi_{\rm{gs}}}$. Dashed lines indicate the ballistic light-cone velocity $v=J=1$. The wrap-around features at $t>N/2$ reflect the periodic boundary conditions.
    (b,c) Local magnetization $S^z_i(t)$ at the quench site (b) $i=0$ and the quarter-chain site (c) $i=16$.
    (d,e) Real (d) and imaginary (e) parts of the unequal-time spin correlator $C^{-+}_{16,0}(t)$.
    (f) Normalized dynamical structure factor $\tilde{S}^{-+}(k,\omega) = S^{-+}(k,\omega)/\max_{\omega}S^{-+}(k,\omega)$ at $k=n\pi/4$ ($n=0,1,2,3,4$). Curves are vertically offset for clarity.
    All rot-tVMC results are obtained with a $D=16$ periodic MPS ansatz. In panels (b)--(f), exact references are shown in black.
    }\label{fig:N64_XY_benchmark}
\end{figure*}

\section{Rotation Sampling for local-operator Quenches}\label{sec:rotation_realtime}

We next turn to the charged local-operator quench, a paradigmatic case of support mismatch. Here a charged local operator such as the raising operator $S^+$ enforces an up spin at the quench site $i$, so every configuration $\ket{\sigma}$ with a down spin at $i$ has zero amplitude at $t=0$ and is never sampled from $|\psi(\sigma)|^2$, causing the local support mismatch confined to the quench site $i$.

\subsection{Unbiasing time-dependent variational Monte Carlo with rotation sampling}\label{sec:unbias_tVMC}

To mitigate the local support mismatch, we introduce a one-site unitary $U$ acting on the quench site $i$, and augment the variational state by a parameter-independent helper state $\ket{\psi_{\rm h}}$ defined in an orthogonal symmetry sector associated with the local rotation $U$.
The resulting rotated representation of the state and Hamiltonian takes the form
\begin{equation}
  \ket{\tilde\Psi}
  =
  U\left(\ket{\psi}+c_h\ket{\psi_{\rm h}}\right),
  \quad
  \tilde H=UHU^\dagger .
\end{equation}
In the rotated representation, sampling is performed from the Born distribution $|\tilde{\Psi}(\sigma)|^2$, where
\begin{equation}
  \tilde{\Psi}(\sigma)
  \equiv \braket{\sigma|\tilde\Psi}
  =
  \sum_{\sigma'} U_{\sigma\sigma'}
  \left[\psi(\sigma')+c_h\,\psi_{\rm h}(\sigma')\right].
\end{equation}
Configurations that had zero amplitude in the original basis can then acquire finite sampling weight through the helper component and contribute to the estimators. Because the Hamiltonian preserves the symmetry and the helper state lies in an orthogonal symmetry sector, the helper state does not couple to the dynamics of the variational state. Therefore, rotation sampling with an augmented helper wave function eliminates the support mismatch without altering the real-time dynamics. The detailed analysis is provided in the Appendix~\ref{sec:methods_detail}.

Figure~\ref{fig:4x4_tVMC_benchamrk} compares standard tVMC and \ac{rot-tVMC} for the $4\times4$ periodic square-lattice XY model. We use a \ac{RBM}~\cite{Carleo2017} with hidden-unit density $\alpha=M/N=4$ and no bias terms. Standard tVMC remains frozen at the initial value of $S^z_0(t)$, while \ac{rot-tVMC} fixes the support-mismatch problem and follows the exact reference over the simulated time window. 

To assess whether \ac{rot-tVMC} can accurately capture long-time dynamics, we consider a local quench in a periodic spin-$1/2$ XY chain with $N=64$ sites, which is exactly solvable via the Jordan--Wigner transformation~\cite{Lieb1961SolubleAFChain,Jordan1928}. We use a periodic \ac{MPS}~\cite{PerezGarcia2007MPS, xiangDensityMatrixTensor2023} to match the ring geometry, with bond dimension $D=16$. We first optimize it in the $S^z_{\rm tot}=0$ sector to obtain the ground state $\ket{\psi_{\rm gs}}$, and then initialize the local quench by applying $S^+_0$ directly on this \ac{MPS} at site $i=0$, yielding $\ket{\psi_0} = S^+_0 \ket{\psi_{\rm gs}}$. 

Figure~\ref{fig:N64_XY_benchmark}(a) shows the space--time profile of $S^z_i(t)$. Following the local spin flip, two wavefronts propagate ballistically away from the quench site with light-cone velocity $v=J=1$, indicated by the dashed lines~\cite{Lieb1972}, and the \ac{rot-tVMC} signal faithfully reproduces this light-cone structure. At later times, the wavefronts enter from the opposite side, leading to wrap-around features that become visible for $t > N/2$.

\begin{figure}[t]
  \centering
  \includegraphics[]{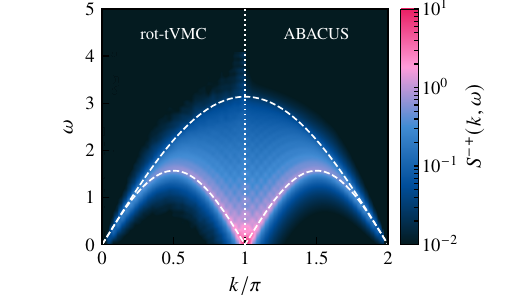}
  \caption{Dynamical structure factor $S^{-+}(k,\omega)$ of the $N=64$ periodic Heisenberg chain. The vertical dotted line at $k=\pi$ separates the two halves. The left half $k\in[0,\pi]$ shows results from rot-tVMC, while the right half $k\in[\pi,2\pi]$ shows the ABACUS result as a reference. The rot-tVMC result is obtained by Fourier transforming real-time correlators, computed with a $D=16$ periodic MPS ansatz, up to $t_{\max}=40$. Dashed lines mark the lower and upper boundaries of the two-spinon continuum.
  }\label{fig:heisenberg_dsf}
\end{figure}

\begin{figure}[t]
  \centering
  \includegraphics[]{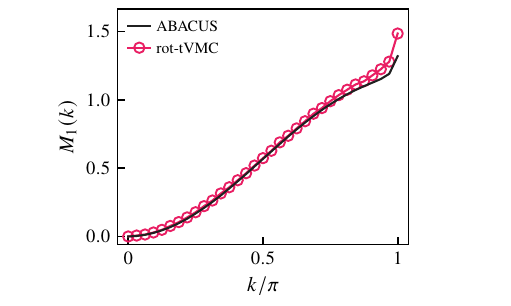}
  \caption{First spectral moment $M_1(k)$ [Eq.~(\ref{eq:sumrule_M1})] of the $N=64$ periodic Heisenberg chain, comparing rot-tVMC with the ABACUS reference. 
  }\label{fig:heisenberg_dsf_M1}
\end{figure}

Figures~\ref{fig:N64_XY_benchmark}(b) and (c) show the local magnetization $S^z_i(t)$ at two representative sites, the quench site $i=0$ and the quarter-chain site $i=16$. At the quench site, $S^z_0(t)$ starts from $1/2$ and then oscillates as the localized excitation disperses into the chain. At the quarter-chain site, $S^z_{16}(t)$ remains close to zero before the wavefront arrives and then begins to oscillate. For both sites, the \ac{rot-tVMC} results track the exact reference up to $t_{\rm max}=40$.

A more comprehensive benchmark is the \ac{DSF}, a quantity accessible in spectroscopic experiments. Since extracting the \ac{DSF} requires unequal-time spin correlations,
\begin{equation}
C^{-+}_{j,0}(t)\equiv\braket{\psi_{\mathrm{gs}}|e^{iHt}S_j^- e^{-iHt} S_0^+|\psi_{\mathrm{gs}}},
\end{equation}
we first present $C^{-+}_{16,0}(t)$ in Figs.~\ref{fig:N64_XY_benchmark}(d) and (e), showing its real and imaginary parts, respectively. This correlator provides a more stringent test, as it is sensitive to phase accumulation during real-time evolution. 
\ac{rot-tVMC} accurately follows the exact oscillatory behavior over most of the simulated time window. Slight deviations appear in the imaginary part for $t>20$, but these remain very small and are on the order of sampling noise.

By Fourier transforming the correlators $C_{j,0}^{-+}(t)$ in space and time, we obtain the \ac{DSF} $S^{-+}(k,\omega)$. Figure~\ref{fig:N64_XY_benchmark}(f) shows the normalized spectral cuts $\tilde{S}^{-+}(k,\omega)=S^{-+}(k,\omega)/\max_\omega S^{-+}(k,\omega)$ at five representative momenta $k=n\pi/4$ ($n=0,1,2,3,4$). The \ac{rot-tVMC} spectra reproduce the main peak positions and line shapes across these momenta, in good agreement with the exact results. The \ac{rot-tVMC} and exact correlators are processed using the same Fourier transform procedure with a finite-time window ($t_{\rm max} = 40$), Lorentzian broadening ($\eta = 0.1$), and a zero-padding factor ($N_{\rm padding} = 4$), ensuring a fair comparison of real-time dynamics.

Having verified the protocol in the XY chain, we next apply it to the periodic spin-$1/2$ Heisenberg chain for a further benchmark. Its finite-size spectrum can be accessed via the algebraic Bethe ansatz~\cite{caux2005ComputeDSF,caux2005anisotropicHeisenberg,caux2009ABACUS}, providing an independent reference beyond the XY case. Figure~\ref{fig:heisenberg_dsf} compares the \ac{rot-tVMC} spectrum with reference results obtained using ABACUS~\cite{caux2009ABACUS}, a numerical implementation based on the algebraic Bethe ansatz. The \ac{rot-tVMC} spectrum is obtained using the same variational ansatz and Fourier-transform parameters as in the XY case above. The two results agree closely and correctly reproduce the lower and upper boundaries of the two-spinon continuum~\cite{muller1981TwoSpinon}. 
Fig.~\ref{fig:heisenberg_dsf_M1} tests the first-moment sum rule
\begin{equation}
    M_1(k)=\int d\omega\, \omega S^{-+}(k,\omega) \label{eq:sumrule_M1}
\end{equation}
on the \ac{rot-tVMC} data against the ABACUS reference. \ac{rot-tVMC} shows good agreement, except for a small deviation near $k=\pi$, likely due to finite-time effects.

Benchmarks on these two one-dimensional chains show that \ac{rot-tVMC} removes the bias in local-operator quench dynamics and enables accurate long-time evolution, allowing the extraction of the \ac{DSF} within the tVMC framework. This extends previous tVMC calculations of spectral functions in settings without support mismatch~\cite{Tiago2023tVMCSpectralFunction} to transverse spectra of Hamiltonians with Abelian symmetries, where a charged local operator such as $S^+$ induces a support mismatch at the initial time.

\subsection{Enhancing projected time-dependent variational Monte Carlo with rotation sampling}\label{sec:enhance_ptVMC}

Extending unbiased local-quench simulations to two dimensions is challenging, as the large parameter space increases Monte Carlo fluctuations, leads to an ill-conditioned and rank-deficient \ac{QGT}~\cite{Park2020}, and amplifies the sampling cost required for accurate simulations~\cite{Markus2020PRL10x10}. \ac{p-tVMC}~\cite{ptvmc1Sinibaldi2023,ptvmc2Gravina2025} provides a systematic route for addressing these challenges. Instead of directly solving the stochastically estimated \ac{TDVP} equation at once, \ac{p-tVMC} updates the state by iteratively minimizing the infidelity
\begin{equation}
    L(\ket{\psi},\ket{\phi})
    =
    1-\frac{|\braket{\psi|\phi}|^2}
    {\braket{\psi|\psi}\braket{\phi|\phi}} ,
\end{equation}
where $\ket{\psi}$ is the variational state to be optimized and $\ket{\phi}$ is a substep target state generated from the current state $\ket{\psi_0}$.
Although slower compared to tVMC, p-tVMC has shown strong numerical stability in 2D global-quench dynamics~\cite{ptvmc2Gravina2025}. 

\begin{figure}[t]
  \centering
  \includegraphics[]{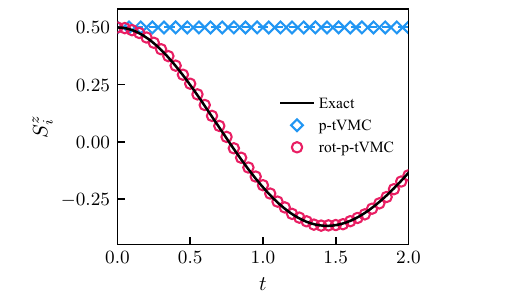}
  \caption{Time evolution of the quench-site magnetization $S^z_{i}(t)$ for the $4\times4$ periodic square-lattice XY model. Standard p-tVMC remains pinned at the initial value, while rot-p-tVMC follows the exact reference. Both simulations use an $\alpha=4$ RBM ansatz, LPE-3, and $50$ optimization steps per stage.
  }\label{fig:benchmark_RBM_ptvmc_vs_ED}
\end{figure}

However, extra care is still needed for local-operator quenches. Figure~\ref{fig:benchmark_RBM_ptvmc_vs_ED} shows this issue for a standard \ac{p-tVMC} simulation based on a third-order \ac{LPE}, using an $\alpha=4$ \ac{RBM} ansatz and $50$ optimization steps per stage. The quench-site magnetization remains pinned at $0.5$ instead of following the exact reference. The origin can be traced to how the short-time evolution is represented. In the \ac{LPE} update, the states entering the infidelity minimization are
\begin{equation}
    \ket{\psi}=\ket{\psi_0},\quad \ket{\phi}=(1+a_i\tau H)\ket{\psi_0},
\end{equation}
where $a_i$ are the \ac{LPE} coefficients and $\ket{\psi_0}$ is the state at the beginning of the substep. For the local $S^+$ quench, $\ket{\psi_0}$ initially has zero amplitude on configurations with a down spin at the quench site. The minimization therefore lacks sampling support on these configurations. This biases the infidelity gradient and can make the optimization frozen at the initial state.

\begin{figure*}[tb]
  \centering
  \includegraphics[width=0.98\textwidth]{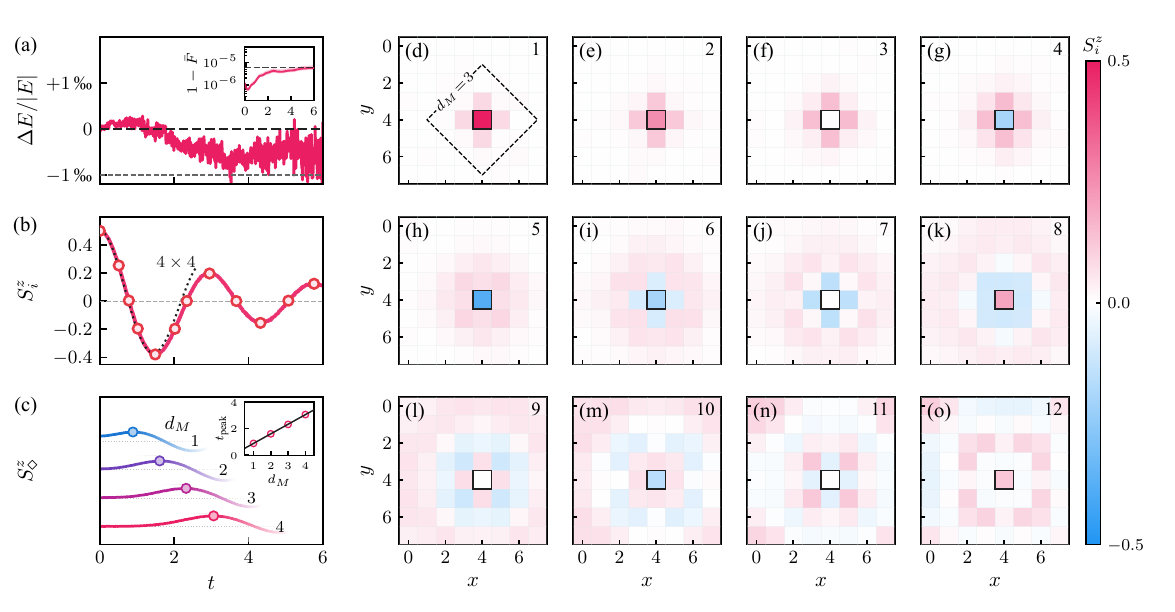}
  \caption{Local-operator-quench dynamics of the $8\times8$ periodic square-lattice XY model.
  (a) Relative energy drift $\Delta E/|E|$. The inset shows the average infidelity $1-\overline{F}$. 
  The shaded band gives the minimum--maximum range over the three substeps at each time, and the dashed line marks $6\times10^{-6}$.
  (b) Quench-site magnetization $S_{i}^z(t)$. The exact result for a $4\times4$ lattice is shown as a reference for the early-time quench dynamics, and the circle markers indicate the times used for the spatial snapshots in panels (d)--(o). 
  (c) Magnetization summed over diamond shells with fixed Manhattan distance $d_M$ from the quench site. The inset plots the peak time $t_{\rm peak}$ versus $d_M$. The linear fit is shown as a guide to ballistic light-cone propagation.
  (d)--(o) Spatial snapshots of the on-site magnetization $S_i^z$ after the quench. The quench site $i=(4,4)$ is outlined by a square box, and the dashed diamond in panel (d) illustrates the $d_M=3$ shell.
  All rot-p-tVMC results are obtained with an $\alpha=8$ RBM ansatz.
  }\label{fig:2D_8x8}
\end{figure*}

This issue may be mitigated by modifying the initial state. However, using a random initial state shifts the task from dynamics simulation to state search, which is more challenging and time-consuming~\cite{ptvmc2Gravina2025}. Rotation sampling provides a systematic way to remove this support bias. 
We perform the infidelity minimization in a rotated representation generated by a two-site unitary acting on the quench site and a neighboring site. This unitary is chosen to conserve $S^z_{\rm tot}$, so that the rotated representation preserves the $U(1)$ symmetry of the Hamiltonian. In this representation, previously inaccessible regions of the configuration space acquire finite sampling weight, allowing the geometric and gradient information required for the update to be captured by the Monte Carlo estimation. 
It is worth noting that, under the rotation, configurations such as $\ket{\downarrow \downarrow \cdots}$ remain zero-amplitude. Nevertheless, unlike tVMC, the present scheme is formulated as an optimization problem where strictly unbiased estimators at every step are not required, provided that the optimization does not become totally frozen.
As shown in Fig.~\ref{fig:benchmark_RBM_ptvmc_vs_ED}, the resulting \ac{rot-p-tVMC} removes the freezing and follows the exact reference. This integration therefore combines the stability of \ac{p-tVMC} with the improved sampling support provided by rotation sampling.

Figure~\ref{fig:2D_8x8} presents the \ac{rot-p-tVMC} simulation of the 2D XY model on an $8\times8$ periodic square lattice, using an $\alpha=8$ \ac{RBM} ansatz with $32768$ complex parameters. Fig.~\ref{fig:2D_8x8}(a) shows the relative energy drift during the evolution. The drift remains below $1\,\text{\textperthousand}$ over the full time window, even though statistical fluctuations increase at later times. The inset shows the average infidelity during the \ac{rot-p-tVMC} evolution, which remains small, below $6\times10^{-6}$. These diagnostics demonstrate the numerical stability and accuracy of \ac{rot-p-tVMC}.

Figure~\ref{fig:2D_8x8}(b) shows the quench-site magnetization, and panels (d)--(o) show spatial snapshots of $S^z(x,y;t)$. At early times, the \ac{rot-p-tVMC} result agrees with the exact $4\times4$ reference because the propagating wavefront has not yet reached the boundary of the $4\times4$ system. The two curves begin to deviate once the wavefront has propagated beyond the smaller-system length scale, reflecting the finite-size effect.

The spatial snapshots [Fig.~\ref{fig:2D_8x8}(d-o)], taken at the times marked by circles in (b), show ballistic spreading organized by the square-lattice geometry. For nearest-neighbor couplings, the natural distance from the quench site $i$ is the shortest bond-path length, i.e., the Manhattan distance $D_M(i,j)$. The wavefronts therefore form diamond-shaped shells of constant $d_M=D_M(i,j)$~\cite{Carleo2014LightCone}, and the case $d_M=3$ is illustrated by the dashed box in Fig.~\ref{fig:2D_8x8}(d).
To quantify this spreading, Fig.~\ref{fig:2D_8x8}(c) shows the magnetization summed over a diamond shell,
\begin{equation}
    S^z_{\Diamond}(d_M,t) =
    \sum_{j:\,D_M(i,j)=d_M} S_j^z(t),
\end{equation}
where the sum is taken over all sites at Manhattan distance $d_M$ from the quench site $i$.
For each $d_M$, we extract the peak time $t_{\mathrm{peak}}(d_M)$ of $S^z_{\Diamond}(d_M,t)$. The inset shows a clear linear dependence of $t_{\mathrm{peak}}$ on $d_M$, consistent with ballistic light-cone propagation after the quench.

\section{Rotation Sampling for Ground-State Optimization}\label{sec:rotation_gs}

The support-mismatch problem is not unique to real-time evolution; it also arises in ground-state optimization, where learning nontrivial sign structures without prior knowledge is particularly difficult~\cite{liang2018FrustratedSignCNN,Choo20192DJ1J2NQS,Westerhout2020GeneralizationPropertiesofNeuralNetwork,Chen2022PRRSolvingSignStructure}. This difficulty can be partly attributed to the fact that refining a nontrivial sign structure can drive wave-function amplitudes through zero. Near wave-function nodes, the gradient signal-to-noise ratio deteriorates sharply, leading to unstable or stalled optimization~\cite{misery2025lookingelsewhereimprovingvariational}. Basis rotation can therefore also benefit ground-state optimization.

However, unlike the charged local-operator quenches discussed above, where the lost support is localized at the quench site and can be targeted directly, the missing support in ground-state optimization is generally unknown. In VMC, ground-state optimization is typically performed using \ac{SR}~\cite{Sorella1998SR1,Sorella2001SR2,Sorella2005SR3}. We therefore augment SR with Monte Carlo sampling in randomly rotated bases, a scheme we refer to as rotSR. During optimization, we apply a random two-site rotation that conserves $S^z_{\mathrm{tot}}$ to each bond in turn. The variational update is evaluated in the corresponding rotated basis, partially allowing configurations suppressed in the original basis to acquire finite sampling weight and contribute to the update. 
Rotation sampling leaves the asymptotic complexity of standard SR unchanged, while increasing the cost of each wave-function evaluation by a factor of $N_{\mathrm{conn}}$, the number of basis configurations connected by the two-site unitary. For the spin-$1/2$ systems considered here, $N_{\mathrm{conn}}=1$ or $2$, resulting in only a small constant-factor overhead.

\begin{figure}[t]
  \centering
  \includegraphics[]{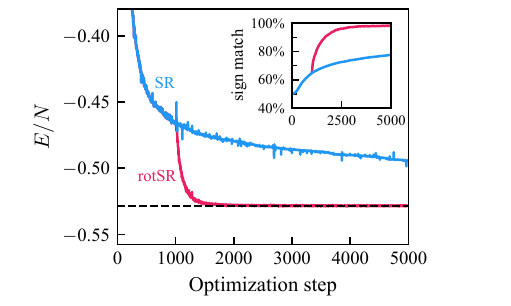}
  \caption{Energy optimization of the $4\times4$ periodic square-lattice $J_1$-$J_2$ model with a full lookup-table wavefunction; the corresponding sign match is shown in the inset. The dashed line marks the exact ground-state energy. At step $1000$, rotation sampling is applied, after which the rotSR energy resumes a rapid decrease toward the exact value and the sign match improves to $100\%$.}
  \label{fig:rotsampling_gs}
\end{figure}

Figure~\ref{fig:rotsampling_gs} compares the two optimization methods for the ground-state energy optimization of the $4\times4$ periodic square-lattice $J_1$–$J_2$ model at the frustrated point $J_2/J_1 = 0.5$, with the sign match during optimization shown in the inset. 
To isolate sampling effects from the expressivity limitations of the ansatz and optimization difficulties associated with a complex energy landscape, we use a full lookup-table wave function with $C_{16}^8=12870$ independent parameters. Unlike the unfrustrated Heisenberg model, whose ground state obeys the Marshall sign rule~\cite{marshall1955HeisenbergMarshallSign}, the $J_1$-$J_2$ model exhibits a highly nontrivial sign structure~\cite{liang2018FrustratedSignCNN,Chen2022PRRSolvingSignStructure,Choo20192DJ1J2NQS,Westerhout2020GeneralizationPropertiesofNeuralNetwork}. Its optimization is therefore demanding and typically requires substantial computational resources to reach high accuracy~\cite{fan2026disentanglingtensornetworkstates}. 

Our results indicate that, in addition to the difficulty of the optimization landscape associated with a complex ansatz manifold, Monte Carlo sampling itself can impose an additional limitation. Standard \ac{SR} initially shows rapid convergence. However, as the optimization proceeds, the wave function develops sharp spikes, indicating increased variance and suggesting that some amplitudes are being driven toward zero as the wave function attempts to change their signs. To quantify this, we track the \emph{sign match}, the fraction of configurations whose amplitude sign agrees with that of the exact ground state (see App.~\ref{app:sign_match} for the explicit definition). As the energy optimization slows, the sign match ceases to improve. When the sampling is switched at step $1000$ to a randomly chosen frame constructed from local two-site rotations, the ground-state energy resumes a rapid decrease and approaches the exact ground-state energy, while the sign match is further refined towards $100\%$.

\section{Conclusion and Discussion}\label{sec:discussion}
In this work, we leverage basis flexibility to resolve the support-mismatch problem in variational Monte Carlo, a technique referred to as rotation sampling. Operating exclusively at the sampling stage, this approach is highly modular and readily compatible with existing algorithms. 
Integrated with tVMC, it provably removes the structural sampling bias from charged local-operator quenches while leaving the physical dynamics unchanged, enabling accurate long-time evolution and the extraction of transverse dynamical structure factors in one-dimensional spin chains. Combined with p-tVMC, it eliminates the dynamical freezing and enables stable two-dimensional local-quench simulations, capturing the ballistic spreading of a local excitation. Beyond real-time dynamics, sampling in randomly rotated local bases also accelerates ground-state optimization within SR.

Several directions remain to be explored. Adaptive or optimized rotations may further reduce sampling variance and improve convergence. Larger local rotations, symmetry-adapted transformations, or classically simulable quantum circuits could extend the accessible sampling bases. Such extensions may also be relevant for variational quantum algorithms, where basis rotations can be implemented directly in quantum circuits.

More broadly, our results suggest that the freedom of representation can itself be used as a sampling resource, the same freedom that makes computational difficulty basis dependent. Beyond the settings studied here, this perspective may prove useful in condensed matter physics, statistical physics, quantum chemistry, and deep-learning based methods.

\begin{acknowledgments}
We thank Jiajun Yu, Kang Wang, and Jannes Nys for helpful discussions. This work was supported by the National Natural Science Foundation of China (Grant No. 12488201), the National Key Research and Development Program of China (2021ZD0301800). 
Y.W. is supported by a start-up grant from
IOP-CAS.

\end{acknowledgments}

\section*{Data availability}
The data supporting this study will be made openly available upon publication.

\section*{Code availability}
The implementation of rotation sampling used in this work is developed on top of NetKet~\cite{netket2:2019,netket3:2022} and ptvmc-systematic-study package~\cite{netket_fidelity}. The code repository will be made publicly available upon publication.

\appendix

\section{Rotation-Sampling tVMC}\label{sec:methods_detail}

 Here we describe the rotation-sampling scheme for real-time evolution. For a variational state $\ket{\psi(\bm{\theta})}$, the parameter update is obtained by minimizing
\begin{equation}\label{eq:norm_dist}
  \mathcal{D}(\mathrm{d}\bm{\theta})
  =
  \left\lVert
    \ket{\psi(\bm{\theta}+\mathrm{d}\bm{\theta})}
    -
    \mathrm{e}^{-\mathrm{i}H\mathrm{d}t}
    \ket{\psi(\bm{\theta})}
  \right\rVert_2 .
\end{equation}
Linearizing this expression gives the uncentered \ac{SR} equation
\begin{equation}\label{eq:sr_unc}
  S^{\mathrm{unc}}_{kk'}\,\mathrm{d}\theta_{k'}
  =
  -\mathrm{i}\,\mathrm{d}t\,F^{\mathrm{unc}}_k ,
\end{equation}
with
\begin{equation}
  S^{\mathrm{unc}}_{kk'}
  =
  \frac{\braket{\partial_{\theta_k}\psi|\partial_{\theta_{k'}}\psi}}
       {\braket{\psi|\psi}},
  \quad
  F^{\mathrm{unc}}_k
  =
  \frac{\braket{\partial_{\theta_k}\psi|H|\psi}}
       {\braket{\psi|\psi}} .
\end{equation}
Unlike the centered metric commonly used for ground-state optimization, which projects out the component along the state itself, the uncentered form retains it. This component carries the global phase, which is a gauge degree of freedom for ground states but is essential for the unequal-time correlators from which dynamical structure factors are extracted. 

In standard \ac{VMC}, these quantities are estimated by
\begin{equation}\label{eq:rot_unc_estimators}
  S^{\mathrm{unc}}_{kk'}
  = \left\langle O_k^*  O_{k'} \right\rangle_{P},
  \qquad
  F^{\mathrm{unc}}_k
  = \left\langle O_k^* E_{\mathrm{loc}} \right\rangle_{P},
\end{equation}
where
\begin{equation}\label{eq:local_estimators}
  E_{\mathrm{loc}}(\sigma) 
  = \frac{\langle \sigma|H|\psi\rangle}
         {\braket{\sigma|\psi}},
  \qquad
  O_k(\sigma) 
  = \frac{\braket{\sigma|\partial_{\theta_k}\psi}}
         {\braket{\sigma|\psi}}.
\end{equation}
with samples are drawn from
\begin{equation}
    P(\sigma)= \frac{ |\braket{\sigma|\psi}|^2 }{\braket{\psi|\psi} }.
\end{equation}

For local-operator quenches, this distribution contains structural zero-amplitude regions, as discussed above. Omitting the affected configurations biases the estimators. Rotation sampling removes this bias by sampling in a rotated representation, built from two ingredients. First, to accommodate a local quench such as a spin flip at site $i$, we apply a general one-site (non-$S^z_{\mathrm{tot}}$-conserving) unitary $U$ at $i$. Because $U$ does not conserve $S^z_{\mathrm{tot}}$, it mixes the zero-amplitude configurations with configurations from other $S^z_{\mathrm{tot}}$ sectors. The specific choice of $U$ is not critical, and in practice we use a Hadamard-like rotation that mixes the spin-up and spin-down states at $i$. Rotation alone is not sufficient, however, since those sectors are themselves empty for the physical state. We therefore add a second ingredient, a parameter-independent helper state $\ket{\psi_{\rm h}}$ residing in an orthogonal $S^z_{\mathrm{tot}}$ sector, yielding the augmented state
\begin{equation}
  \ket{\Psi} = \ket{\psi} + c_h\ket{\psi_{\rm h}}.
\end{equation}
The helper supplies finite amplitude in a neighboring $S^z_{\mathrm{tot}}$ sector, so that $U$ maps the previously zero-amplitude configurations into it and they acquire finite sampling weight. Concretely, the $S^+_i$ quench studied here acts on a ground state in the $S^z_{\mathrm{tot}}=0$ sector, so the physical state sits in the $S^z_{\mathrm{tot}}=1$ sector and the helper in the $S^z_{\mathrm{tot}}=2$ sector. Sampling is then performed in the rotated representation
\begin{equation}
  \ket{\tilde\Psi}=U\ket{\Psi},\qquad \tilde H=UHU^\dagger ,
\end{equation}
with configurations drawn from the rotated distribution
\begin{equation}\label{eq:P_rot}
  \tilde P(\sigma)
  =\frac{|\braket{\sigma|\tilde\Psi}|^2}{\braket{\tilde\Psi|\tilde\Psi}}
  =\frac{|\braket{\sigma|U|\Psi}|^2}{\braket{\Psi|\Psi}},
\end{equation}
and the corresponding rotated local estimators
\begin{equation}\label{eq:rot_local_estimators}
  \tilde E_{\mathrm{loc}}(\sigma)
  = \frac{\braket{\sigma|\tilde H|\tilde\Psi}}
         {\braket{\sigma|\tilde\Psi}},
  \qquad
  \tilde O_k(\sigma)
  = \frac{\braket{\sigma|\partial_{\theta_k}\tilde\Psi}}
         {\braket{\sigma|\tilde\Psi}}.
\end{equation}
The uncentered \ac{QGT} and force are then estimated as
\begin{equation}
  S^{\mathrm{unc}}_{kk'}=\langle \tilde O_k^*\,\tilde O_{k'}\rangle_{\tilde P}, \quad F^{\mathrm{unc}}_k=\langle \tilde O_k^*\,\tilde E_{\mathrm{loc}}\rangle_{\tilde P}.
\end{equation}

The helper state likewise leaves the physical update unchanged. Since $\ket{\psi_{\rm h}}$ is parameter-independent, $\partial_{\theta_k}\ket{\Psi}=\partial_{\theta_k}\ket{\psi}$, and since the Hamiltonian conserves $S^z_{\mathrm{tot}}$ while $\ket{\psi_{\rm h}}$ lies in an orthogonal sector, all cross terms between the two sectors vanish. Consequently the uncentered \ac{QGT} and force computed from $\ket{\Psi}$ reduce to those of $\ket{\psi}$ up to a common scalar,
\begin{equation}\label{eq:helper_scaling}
  S^{\mathrm{unc},\Psi}_{kk'} = r\,S^{\mathrm{unc},\psi}_{kk'},\qquad
  F^{\mathrm{unc},\Psi}_{k} = r\,F^{\mathrm{unc},\psi}_{k},
\end{equation}
with $r=\braket{\psi|\psi}/\braket{\Psi|\Psi}$. This common factor $r$ multiplies both sides of the uncentered \ac{SR} equation~\eqref{eq:sr_unc} and cancels exactly, so $r$ never needs to be evaluated in practice. The helper therefore alters only the sampling support, leaving the physical parameter update identical to that of $\ket{\psi}$ alone. 

 Because the rotation only needs to provide support for the missing spin-down-at-$i$ configurations, it is enough for $\ket{\psi_{\rm h}}$ to carry amplitude only on configurations with spin up at $i$, and a constant-amplitude helper on that subspace already restores the missing support.  Evaluating the helper component adds negligible cost. Its amplitudes correspond to a static state retrieved from a lookup table, and each step still requires only a single evaluation of the variational ansatz in the physical sector. The coefficient $c_h$ controls how much sampling weight is shifted into the helper sector, and we tune it so that the helper component $c_h\ket{\psi_{\rm h}}$ is a small but non-negligible fraction of the physical wavefunction $\ket{\psi}$, roughly ten percent. 

\begin{figure}[t]
  \centering
  \includegraphics[]{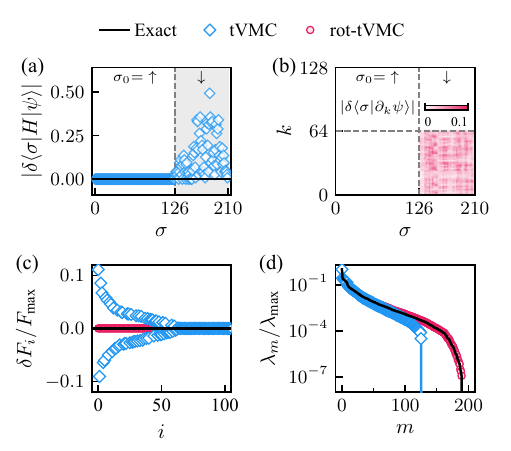}
  \caption{Full-summation diagnostics of estimator bias.
  Diagnostics are performed for the local-quench initial state $S^+_0\ket{\psi_{\rm gs}}$ of the $N=10$ periodic XY chain, represented by a periodic MPS ansatz with bond dimension $D=8$.
  (a) Deviation of the energy contribution, $|\delta\braket{\sigma|H|\psi}|$, shown as a function of the configuration index $\sigma$.
  (b) Deviation of the gradient contribution, $|\delta\braket{\sigma|\partial_k\psi}|$, shown as a function of the configuration index $\sigma$ and parameter index $k$.
  (c) Sorted relative deviations of the force-vector components, $\delta F_i/F_{\max}$.
  (d) Normalized eigenvalue spectrum, $\lambda_m/\lambda_{\max}$, of the uncentered QGT.
  The vertical dashed lines in (a) and (b) separate configurations with $\sigma_0=\uparrow$ from those with $\sigma_0=\downarrow$. The horizontal dashed line in (b) separates the two MPS local-tensor parameter blocks, corresponding to physical indices $s=\uparrow$ and $s=\downarrow$. The force-vector deviations in (c) and the QGT spectrum in (d) are normalized by $F_{\max}$ and $\lambda_{\max}$, respectively.}
  \label{fig:qgt_diagnostics}
\end{figure}

 Finally, Fig.~\ref{fig:qgt_diagnostics} shows a full-summation diagnostic of the bias for an N = 10 XY chain after an $S^+_0$ excitation of the ground state, represented by a $D=8$ periodic \ac{MPS} ansatz. Because the \ac{MPS} parameters are not shared between sites or onsite spin sectors, this setup allows a detailed localization of the missing contributions in both configuration space and parameter space. 

For zero-amplitude configurations, the local estimators $E_{\rm loc}(\sigma)$ and $O_k(\sigma)$ are singular. However, the amplitude-weighted quantities $\braket{\sigma|H|\psi}$ and $\braket{\sigma|\partial_k\psi}$ remain finite and contribute to the exact force vector $F$ and \ac{QGT} $S$. Figure~\ref{fig:qgt_diagnostics}(a) shows that the omitted energy contribution appears only for configurations with a down spin at the quenched site. Figure~\ref{fig:qgt_diagnostics}(b) shows the corresponding gradient contribution. Since the \ac{MPS} parameters are not shared between the two onsite spin sectors, the gradient bias can also be resolved in the parameter space. We find that it is localized on the local tensor at the quench site and confined to the parameter block associated with the down-spin physical index. Figures~\ref{fig:qgt_diagnostics}(c) and (d) show the resulting errors in the force vector $F$ and \ac{QGT} $S$. Standard sampling gives component-wise force-vector errors and a rank-deficient \ac{QGT}. Rotation sampling instead agrees with the exact full-summation result within numerical precision and recovers the full \ac{QGT} spectrum.

\section{Calculation Parameters}\label{sec:calc_details}
In all \ac{rot-tVMC} simulations, we use $40960$ samples to evaluate these estimators. The \ac{TDVP} equations are integrated using a fourth-order Runge-Kutta scheme~\cite{Butcher2016} with a time step of $\delta t = 0.01$. A canonical diagonal shift $\lambda=10^{-8}$ is applied to stabilize the uncentered \ac{QGT}.

In the \ac{rot-p-tVMC} simulation of the $8 \times 8$ periodic square lattice XY model, we use $4096$ samples with the LPE-3 scheme, with a time step of $\delta t = 0.01$. Each optimization stage comprised 500--1000 iterations, with learning rates and automatic damping tuned within the ranges of 0.01--0.05 and $10^{-8}$--$10^{-2}$, respectively.

In both simulations, we deactivate the rotation sampling once the previously zero-amplitude configurations have acquired enough sampling probability to be sampled reliably, which in practice happens after roughly $20$ time steps. We then switch to standard sampling for the subsequent evolution. 

For the ground-state optimization of the $4\times4$ periodic square-lattice $J_1$-$J_2$ Heisenberg model, we use $4096$ samples and $5000$ optimization steps. The optimization uses a learning rate of $0.05$ and a diagonal shift $\lambda = 0.005$.

\section{Initialization of Local Excitations}

We initialize locally excited neural quantum states by applying a learnable single-site operator to a reference ground state $\psi_{\mathrm{base}}(\sigma)$,
\begin{equation}
\psi_{\mathrm{exc}}(\sigma)
= P_{S_z}\sum_{k \in \{\uparrow,\downarrow\}} M_{\sigma_i, k} \,
\psi_{\mathrm{base}}(\sigma^k_i),
\end{equation}
where $P_{S_z}$ is a projection operator that maps the wavefunction to the target excited-state sector, $\sigma^{k}_i$ denotes the configuration with the spin at site $i$ fixed to $k$, and $M \in \mathbb{C}^{2\times 2}$ is a variational matrix. In particular, we choose
\begin{equation}
M =
\begin{pmatrix}
\varepsilon & 1 \\
\varepsilon & \varepsilon
\end{pmatrix}.
\end{equation}
When $\varepsilon=0$, this strictly realizes the action of the operator $S^+_i$. For numerical stability, in practice we use a negligible perturbation $\varepsilon=10^{-20}$ to avoid tangent-space collapse. Allowing $M$ to be variational provides a convenient initialization for the subsequent dynamics.

\section{Calculation of the Dynamical Structure Factor}

To compute the dynamical structure factor, we first calculate the time-dependent spin correlator,
\begin{eqnarray}\label{eq:Cmp_methods}
  C^{-+}_{j,0}(t) &= &\frac{\braket{\psi_{\mathrm{gs}} \vert \mathrm{e}^{iHt} S^-_j \mathrm{e}^{-iHt} S^+_{0} \vert \psi_{\mathrm{gs}}}}{\braket{\psi_{\mathrm{gs}} \vert \psi_{\mathrm{gs}}}} \nonumber \\
  &= &\frac{\braket{\psi_{\mathrm{gs}} \vert S^-_j \mathrm{e}^{-i(H-E_{\mathrm{gs}})t} S^+_{0} \vert \psi_{\mathrm{gs}}}}{\braket{\psi_{\mathrm{gs}} \vert \psi_{\mathrm{gs}}}} .
\end{eqnarray}
By defining the time-evolved excited state as 
\begin{equation}
    \ket{\widetilde\psi(t)} \equiv \mathrm{e}^{-i(H-E_{\mathrm{gs}})t}S_0^+\ket{\psi_{\mathrm{gs}}} ,
\end{equation}
we evaluate the correlator using Monte Carlo sampling as 
\begin{eqnarray}
  C^{-+}_{j,0}(t) &=& \frac{\braket{\psi_{\mathrm{gs}} \vert S^-_j \vert \widetilde\psi(t)}}{\braket{\psi_{\mathrm{gs}} \vert \psi_{\mathrm{gs}}}} \nonumber \\
  &=& \mathcal{N} \sum_x \frac{\vert\braket{x \vert \widetilde\psi(t) } \vert^2}{\braket{\widetilde\psi(t) \vert \widetilde\psi(t)} } \left( \frac{\braket{x \vert S^+_j \vert \psi_{\mathrm{gs}}} }{\braket{x \vert \widetilde\psi(t)}} \right)^*,
\end{eqnarray}
where the prefactor 
\begin{equation}
    \mathcal{N} \equiv \frac{\braket{\widetilde\psi(t) \vert \widetilde\psi(t)} }{ \braket{\psi_{\mathrm{gs}} \vert \psi_{\mathrm{gs}}} } = \frac{\braket{\psi_{\mathrm{gs}} \vert S_0^- S_0^+ \vert \psi_{\mathrm{gs}}} }{ \braket{\psi_{\mathrm{gs}} \vert \psi_{\mathrm{gs}}}}
\end{equation}
is a time-independent constant. The correlator is evaluated by sampling configurations $x$ from the probability distribution 
\begin{equation}
p_t(x) = \frac{ \vert\braket{x \vert \widetilde\psi(t)}\vert^2 }{ \braket{\widetilde\psi(t) \vert \widetilde\psi(t)}}.
\end{equation}
In practice, we apply a single overall rescaling to the MCMC estimates to enforce 
\begin{equation}
    C^{-+}_{0,0}(0) = \frac{\braket{\psi_{\mathrm{gs}} \vert S^-_0 S^+_0 \vert \psi_{\mathrm{gs}}} }{ \braket{\psi_{\mathrm{gs}} \vert \psi_{\mathrm{gs}}} }= \frac{1}{2} ,
\end{equation}
which is a direct consequence of the ground state residing in the $S^z_{\mathrm{tot}}=0$ sector.

The dynamical structure factor is then obtained via the spatial and temporal Fourier transforms of the correlator $C^{-+}_{j,0}(t)$. We first perform the spatial Fourier transform,
\begin{equation}\label{eq:Skt_methods}
  S^{-+}(k_n,t) = \frac{1}{N}\sum_{j=0}^{N-1}
  \mathrm{e}^{-i k_n j}\,C^{-+}_{j,0}(t),
\end{equation}
where $k_n=2\pi n/N$. Subsequently, we perform the temporal Fourier transform,
\begin{eqnarray}
  S^{-+}(k,\omega) &=& \frac{1}{2\pi} \int_{-\infty}^{\infty} \mathrm{d} t \, \mathrm{e}^{i\omega t} S^{-+}(k,t) \nonumber\\
  &=& \frac{1}{\pi} \mathrm{Re} \!\int_0^{\infty}\! \mathrm{d} t \, \mathrm{e}^{i\omega t} S^{-+}(k,t).
\end{eqnarray}
We evaluate this integral by introducing a Lorentzian broadening factor $\eta$,
\begin{equation}\label{eq:Skw_methods}
  S^{-+}(k,\omega)
  = \frac{1}{\pi}\,\mathrm{Re}\!\int_0^{t_{\max}}\!\mathrm{d} t\;
    \mathrm{e}^{i\omega t-\eta t}\,S^{-+}(k,t),
\end{equation}
where $\eta=0.10$ and the maximum simulation time is $t_{\max}=40$. To obtain a denser $\omega$ grid, we employ the zero-padding technique~\cite{press2007NumericalRecipes} with a padding factor of $N_{\rm padding}=4$. No additional window function is applied.

\section{Definition of the sign match}\label{app:sign_match}
To quantify how well the optimized wave function captures the ground-state sign structure, we define the indicator \emph{sign match} as
\begin{equation}
\begin{aligned}
  \mathrm{sign\ match}
  &= \frac{1}{N_{\mathrm{kept}}}
    \sum_{\sigma \in \mathcal{S}_\epsilon}
    \mathbbm{1}\!\left[
      \operatorname{sgn}\psi_\theta(\sigma)
      = \operatorname{sgn}\psi_{\mathrm{ED}}(\sigma)
    \right], \\
  \mathcal{S}_\epsilon &= \{\sigma : |\psi_{\mathrm{ED}}(\sigma)|^2 > \epsilon\}.
\end{aligned}
\end{equation}
Here $\mathbbm{1}[\cdot]$ is the indicator function, equal to $1$ when the enclosed condition holds and $0$ otherwise, so that the sum counts the configurations whose variational sign $\operatorname{sgn}\psi_\theta(\sigma)$ matches that of the exact ground state. We choose $\epsilon = 10^{-8}$ to restrict the comparison to configurations with non-negligible exact weight; $N_{\mathrm{kept}} = |\mathcal{S}_\epsilon|$ is the number of such configurations, each counted with equal weight. The overall sign of $\psi_\theta$ is fixed by choosing the global sign that maximizes the agreement.
\bibliography{references}

\end{document}